\documentclass[letterpaper,twocolumn,10pt]{article}
\usepackage{usenix,epsfig,endnotes}
\usepackage{enumerate}
\usepackage{subfig}
\usepackage{tabularx,booktabs,makecell,multirow,graphicx,amsmath}
\usepackage{xurl}
\usepackage{hyperref}
\usepackage{tablefootnote} 
\usepackage{threeparttable}

\begin{document}

\date{}

\title{\Large \bf OD-MoE: On-Demand Expert Loading for Cacheless Edge-Distributed MoE Inference}

\author{
\rm{Liujianfu Wang}$^{\dag}$ 
\hspace{0.35em} \rm{Yuyang Du}$^{\dag, \ddag}$ 
\hspace{0.35em} \rm{Yuchen Pan} 
\hspace{0.35em} \rm{Soung Chang Liew}$^*$ 
\hspace{0.35em} \rm{Jiacheng Liu} 
\hspace{0.35em} \rm{Kexin Chen} 
\\ 
\\ The Chinese University of Hong Kong 
}

\maketitle

\def\thefootnote{$\dag$}\footnotetext{Equal contribution. $^{\ddag}$Project lead. $^{*}$Corresponding author.}
\renewcommand{\thefootnote}{\arabic{footnote}}

\thispagestyle{empty}

\subsection*{Abstract}
Mixture-of-Experts (MoE), while offering significant advantages as a Large Language Model (LLM) architecture, faces substantial challenges when deployed on low-cost edge devices with tight memory constraints. Expert offloading mitigates this issue by storing expert parameters in CPU memory and caching a subset of popular experts in GPU memory. Although this approach improves GPU memory utilization by caching only the likely-used experts, the GPU memory reserved for expert caching is underutilized compared with dense LLMs. This paper presents \textbf{OD-MoE,} a distributed MoE inference framework that obviates the need for expert caches via fully \underline{\textbf{o}}n-\underline{\textbf{d}}emand expert loading. OD-MoE is built upon two key mechanisms: 1) parallelizing expert loading and expert computation across distributed edge nodes, and 2) an ultra-accurate emulative predictor that forecasts expert activations multiple layers ahead while expert computation is ongoing. With these innovations, OD-MoE dynamically loads each target expert to one of the distributed nodes just-in-time before its activation and promptly evicts it afterward, freeing GPU memory for subsequent experts. We comprehensively benchmark OD-MoE against state-of-the-art MoE offloading systems on a ten-node testbed. Experimental results show that: 1) OD-MoE achieves 99.94\% expert activation prediction accuracy, substantially surpassing all existing methods; and 2) OD-MoE delivers approximately 75\% of the decoding speed of a fully GPU-cached MoE deployment while using only 1/3 of the GPU memory. More importantly, by eliminating the need for expert caches, OD-MoE enables MoE inference on edge nodes with less-than-1GB GPU memory, paving the way for practical MoE deployment of low-cost IoT devices at the edge in the LLM era.

\section{Introduction}\label{Section_I}
The rapid explosion of Large Language Models (LLMs) has led to their widespread application across various fields \cite{chen2025chemminer, du2025llmind, dang2025knowguard}. Beyond deploying LLMs in data centers and accessing them via cloud services, there is a growing demand to bring these models to edge. This shift aims to address challenges such as privacy concerns and the dependence on stable, wide-area network connections, both of which arise from centralized, data-center-based approaches \cite{li2025moe, chen2025slimcaching, xue2024wdmoe}. By moving inference closer to where applications are deployed, edge deployment reduces latency, improves reliability, and alleviates privacy concerns by minimizing the need to transmit sensitive data over wide-area networks. Consequently, efficient LLM inference at the edge has emerged as an active area of research in both academia \cite{zhang2024tinyllama, xue2024powerinfer, chu2024mobilevlm} and industry \cite{apple2024foundationmodels, mishra2023huawei, qualcomm2023stable}.

In recent years, the Mixture of Experts (MoE) architecture has emerged as a significant paradigm for scaling LLMs. Unlike dense models that activate all parameters for every input, MoE models selectively activate a small subset of experts during inference, thus enabling significant model size expansion while maintaining per-request computational efficiency \cite{shahout2025score, liu2024survey}. Yet, MoE models demand a substantial memory footprint, requiring 4-5× more GPU memory than dense models with comparable inference FLOPs \cite{cao2025moe}. This substantial memory requirement has become a major barrier for the deployment of MoE models on resource-constrained edge.

Expert offloading is a technique to address this memory challenge \cite{eliseev2023fast, xue2024moe, tang2024hobbit, yi2023edgemoe, li2025moe, yu2025fmoe, cao2025moe}. This method leverages the sparse activation pattern of MoE by storing the majority of expert parameters in slower yet more abundant CPU memory\footnote{In this paper, CPU memory refers to the DRAM of a device. We do not use SSD/NVMe storage due to their large access delays. We also do not rely on on‑chip CPU caches (L1/L2/L3) because their capacities are negligible compared with an expert’s parameter size.} and loading them into GPU memory only when required for computation. However, due to the limited bandwidth of the CPU-GPU link, loading an expert from CPU to GPU introduces significant latency, especially for edge devices that typically rely on standard PCIe buses rather than high-speed interconnects in data centers.

Existing edge inference systems have attempted to address the I/O bottleneck with various methods. For example, EdgeMoE, HOBBIT, and Mixtral-Offloading \cite{yi2023edgemoe, tang2024hobbit, eliseev2023fast} employ quantization techniques to compress expert parameters; and AdapMoE \cite{zhong2024adapmoe} skips certain experts to alleviate the I/O bottleneck. However, these methods may cause significant degradation in model performance, especially when an important expert is highly compressed or skipped. 

The I/O pressure between CPU and GPU can also be mitigated by caching likely-to-be-used experts in GPU memory. For example, Mixtral-Offloading \cite{eliseev2023fast} chooses to cache most recently used experts, while MoE-Infinity \cite{xue2024moe} caches most frequently used ones. HOBBIT \cite{tang2024hobbit} quantizes experts into different precision levels and prefers keeping high-precision ones in the GPU cache. Beyond GPU memory needed for the ongoing expert computation, these approaches require additional GPU memory for expert caching. The additional memory requirement can potentially limit the edge system’s capacity in hosting large‑scale MoE models.

This paper presents \textbf{\textit{OD-MoE}}, a distributed MoE inference framework that eliminates the need for expert caching achieves by enabling fully \underline{\textbf{O}}n-\underline{\textbf{D}}emand expert loading. \textit{\textbf{OD-MoE achieves 75\% of the decoding speed of an MoE deployment with all experts cached in GPU memory, yet requires only 1/3 of the GPU memory without compromising model performance via methods like quantization or expert skipping.}} We refer readers to Table \ref{tab:benchmark} in Section \ref{Section_IV} for detailed experimental results about the GPU memory requirement and a comprehensive speed evaluation.

Our key contributions, along with the underlying design principles of OD-MoE, are summarized as follows:

Our first contribution is an expert-activation predictor with an accuracy up to 99.94\%, which, to the best of our knowledge, is the highest accuracy reported to date. The predictor is inspired by LLM quantization techniques \cite{polino2018model, gholami2022survey}. With model quantization, the compressed model runs faster and requires less GPU memory, but exhibits highly similar behavior to the original full-precision model, including expert routing. Building on this idea, our method diverges from approaches that predict the next layer’s experts based on the current layer’s state \cite{eliseev2023fast, yi2023edgemoe, yu2025fmoe, xue2024moe, tang2024hobbit}. Instead, we employ a low-cost, quantized MoE model, referred to as the "shadow" model, which runs in parallel with the full-precision model. The shadow model acts as a faster-running emulator to predict the expert activations of the full-precision model several layers ahead. Specifically, this predictor uses the future expert activations that are already unfolded by the scaled-down shadow model to forecast the expert activations of the full model. We refer to this approach as \underline{\textbf{S}}caled \underline{\textbf{E}}mulative \underline{\textbf{P}}rediction \textbf{(\textit{SEP})}.

Our second contribution is the elimination of the need for an expert cache by leveraging the ultra-accurate SEP and parallel expert loading-computation over distributed edge nodes. OD-MoE consists of multiple groups of low-cost edge nodes, each with its own CPU-GPU interconnect. Thanks to the ultra-accurate lookahead predictions provided by SEP, distributed devices are fully aware of the expert activations for subsequent layers. This allows one group of nodes to perform expert computation for the current layer while other groups simultaneously load the experts needed for upcoming layers based on the predicted activations. Through cross-device parallelism in expert loading, OD-MoE loads a target expert to one of the distributed nodes just-in-time before its activation and promptly evicts it after computation is complete, making room for subsequent expert loading into GPU memory. From a prospective of I/O bandwidth utilization, the key enablers of OD-MoE’s cache-free inference are 1) the significantly increased overall CPU-GPU I/O throughput achieved through parallel loading across devices, and 2) minimal I/O bandwidth waste, as SEP’s highly accurate predictions rarely result in incorrect expert loads that trigger reloads.

Our third contribution is the development of two essential alignment mechanisms within SEP: KV cache alignment and token alignment. During the shadow model’s decoding process, it generates new tokens and KV cache autoregressively. Although the shadow model closely mimics the behavior of the full-precision model, differences in precision levels can occasionally result in varying outputs, causing discrepancies in the generated KV caches and tokens. While SEP is highly reliable at the beginning of the inference, its accuracy decreases gradually as discrepancies accumulate in the autoregression process (see Fig. \ref{fig:3_quantized_recalls}). Periodically aligning the shadow model’s tokens and KV cache with those of the full-precision model helps prevent the cumulative propagation of errors from one round to the next. While the idea is straightforward, its implementation requires navigating a nontrivial trade-off between prediction accuracy and alignment cost. Alignment must be performed before the shadow model begins generating the next token, even as the full-precision model’s computation is already underway, resulting in a delayed start for the shadow model. Although the shadow model quickly catches up after alignment, predictions for the first few layers are temporarily unavailable, forcing the inference system to revert to an I/O-bottlenecked state for these layers (see Fig. \ref{fig:5_timeline_with_align}).

Overall, OD-MoE offers several benefits:
\begin{enumerate}
    \item \textbf{Reduced GPU Memory Requirements}: By eliminating the need for an expert cache, OD‑MoE significantly reduces GPU memory usage on edge devices. For instance, in our implementation based on Mixtral‑8×7B \cite{jiang2024mixtral}, the GPU memory footprint per worker node is less than 1 GB, including space for the currently scheduled expert and necessary compute memory. This reduction allows even low‑cost devices, such as Wi‑Fi routers or webcams, to participate in inference tasks, prompting a rethinking of the role of low‑cost, underutilized IoT devices in everyday environments. These devices can now become active contributors to LLM inference systems.
    \item \textbf{Lower Hardware Costs}: OD‑MoE requires only one-third of the GPU memory needed for a fully GPU‑cached deployment, reducing hardware costs by over threefold. As discussed earlier, OD‑MoE leverages low-cost edge GPUs, which are significantly cheaper than data center-grade GPUs due to their smaller memory capacity. This enables cost-effective utilization of entry-level GPUs, which have a lower per-GB memory cost compared to advanced GPUs \cite{coinpoet2023gpu}.
    \item \textbf{Applicability to Data Center Deployments}: While OD-MoE is primarily designed for edge scenarios, its SEP scheme and parallel expert loading mechanism can also benefit data center operations. For instance, accurate predictions of future expert usage can serve as the foundation for on-demand expert replication within a cluster – a proven method for mitigating server workload imbalances \cite{han2025grace}. Additionally, OD‑MoE’s cost-effective approach, which uses less expensive GPUs with smaller memory capacities and efficiently offloads experts without being bottlenecked by GPU I/O, may also be advantageous for data centers.
\end{enumerate}

We have open-sourced OD-MoE. The released resources include project implementations, testing scripts, and a comprehensive benchmarking report that compares our model with previously reported methods for both expert-activation prediction and expert offloading. The report also provides detailed implementations of the baseline methods to ensure reproducibility and fair comparisons. The project is available at \textcolor{blue}{https://github.com/Anonymous/DoubleBlind}.\footnote{The link will be made public upon publication to comply with the double-blind review process.}

\section{Related Work}\label{Section_II}
\subsection{Edge MoE Inference with All Experts Cached}
Distributed inference has emerged as a promising research direction for LLM serving systems utilizing MoE models. The architecture of MoE facilitates model deployment across multiple interconnected nodes, where experts can be deployed and activated according to traffic patterns, network conditions, user demands, and device workloads. Several recent works on distributed MoE inference assume that all experts are pre-loaded into GPU memory, eliminating the need for dynamic expert loading. These works primarily focus on optimizing expert placement to maximize inference throughput. 

For example, CoEL \cite{li2025moe} introduces a collaborative inference framework for expert placement across memory-constrained devices. SlimCaching \cite{chen2025slimcaching} formulates expert placement on edge devices as a knapsack problem (KP) to derive the optimal static caching policy for edge nodes. WDMoE \cite{xue2024wdmoe} considers the influence of wireless channel when placing experts – edge nodes with good channel conditions are preferred when placing frequently used experts since these devices lower inter-node communication delays.

In contrast to these approaches, which assume all expert parameters are pre-cached in GPUs, \textbf{OD-MoE} eliminates the need for any pre-cached experts by relying on dynamic expert loading. This approach significantly reduces GPU memory footprints, enabling practical MoE deployments on low-cost edge devices.

\subsection{Dynamic Expert Loading and I/O Utilization}
As an MoE model activates only a subset of experts during inference and leaves a large portion of experts inactivated, expert offloading offers a cost-effective solution for running MoE models with limited GPU memory. A general principle of expert offloading is storing less popular experts within CPU memory and dynamically loading them into the GPU cache only if they are wanted during model inference. 

However, long expert loading times pose a significant challenge, particularly due to the limited GPU I/O bandwidth on edge devices that rely on PCIe buses for CPU-GPU communication.  To alleviate computation stalls during expert loading, recent works have proposed various expert caching strategies to reduce the volume of dynamic expert loading required. Mixtral-Offloading \cite{eliseev2023fast} and AdapMoE \cite{zhong2024adapmoe} manage expert cache pools by offloading the least recently used (LRU) experts, whereas MoE-Infinity \cite{xue2024moe} evicts the least frequently used (LFU) experts when the expert cache is full. Furthermore, fMoE \cite{yu2025fmoe} incorporates a semantic matching scheme between historical prompt and the current input into the cache management scheme to increase its cache-hit rate. Building on these classical caching strategies, HOBBIT \cite{tang2024hobbit} and EdgeMoE \cite{yi2023edgemoe} further take the LLMs' model-specific features into consideration. For example, HOBBIT, which quantizes experts into different precision levels, prioritizes retaining frequently used high-precision experts when the cache is full. Meanwhile, EdgeMoE favors evicting experts from layers farther away from the one currently being processed.

Another approach to mitigate the computation stalls during expert loading is to sacrifice the model’s performance through expert quantization. This technique, applied in Mixtral-Offloading, AdapMoE, Hobbit, and EdgeMoE \cite{eliseev2023fast, zhong2024adapmoe, tang2024hobbit, yi2023edgemoe}, reduces the volume of data to be loaded. Additionally, AdapMoE \cite{zhong2024adapmoe} introduces a bypass mechanism to skip activating experts that are not cached. 

OD-MoE differs from the above work with its fully on-demand expert loading mechanism, which eliminates the need for any expert cache in GPU memory while maintaining both full model precision and all expert activations.

\subsection{Expert-Activation Prediction}
Accurate expert-activation prediction is the foundation for dynamic expert loading. Prior systems have explored various prediction methods, with two major research directions being investigated. 

The first approach relies on statistical models to predict expert activation. For instance, \cite{yi2023edgemoe, yu2025fmoe} build statistical models based on each expert’s historical expert-activation frequencies (i.e., the popularity of this expert) to predict the most likely expert selection. MoE-Infinity \cite{xue2024moe} further improves this approach by recording each expert’s popularity under different requests – it matches the given prompt to one of those recorded requests and makes expert-activation predictions based on the historical data associated with that request. 

The second approach predicts expert activation on the fly. For example, \cite{eliseev2023fast, zhang2025daop, zhong2024adapmoe} adopt simple heuristic: while the latest embedding vector is fed to the current gating network for expert selection, it is also fed to the subsequent gating network for predicting the next layer’s expert activation. HOBBIT \cite{tang2024hobbit} improves upon this method by aggregating the gating networks of successive layers into a larger gating network with multiple layers so that it can predict the expert activation for successive layers simultaneously.

While these prediction methods have been proven to be effective, an even more precise predictor is highly desirable due to the severe penalty for mispredictions. Experts that are incorrectly preloaded must be evicted and replaced with the correct ones, causing significant system stalls as the planned expert activation must wait for the reloading process to complete.

Our prediction method, \textbf{SEP}, belongs to the on-the-fly prediction category but differs significantly from previous methods. At its core, SEP performs an “emulation” of the full model’s expert activation. Specifically, SEP uses the future expert activations unfolded by a faster shadow model to predict the expert activations of the full-precision model, resulting in naturally more reliable predictions. Moreover, since the shadow model runs faster than the full model, SEP provides predictions several layers ahead of the target expert activation.

With FP16 quantization applied to the shadow model, SEP achieves an average prediction accuracy of 99.94\% across multiple layers. Even when more economical INT8 and NF4 quantization are used, SEP maintains prediction accuracies of 97.34\% and 95.67\%, respectively, outperforming the state-of-the-art prediction accuracies reported in related works.

\section{System Design}\label{Section_III}
Subsection \ref{Section_III_A} presents the distributed architectural design of OD‑MoE, focusing on expert distribution, as well as the scheduling of computations and loading among nodes for the decoding stage. Subsection \ref{Section_III_B} delves into the technical details of SEP alignment operations. Subsection \ref{Section_III_C} describes the operation of the prefilling stage in OD-MoE, which requires a different distribution of expert computations across nodes compared to the decoding stage to achieve optimal performance.

\subsection{OD-MoE Architecture and Decoding-Stage Scheduling}\label{Section_III_A}
\begin{figure}[htbp]
    \centering
    \includegraphics[width=0.88\linewidth]{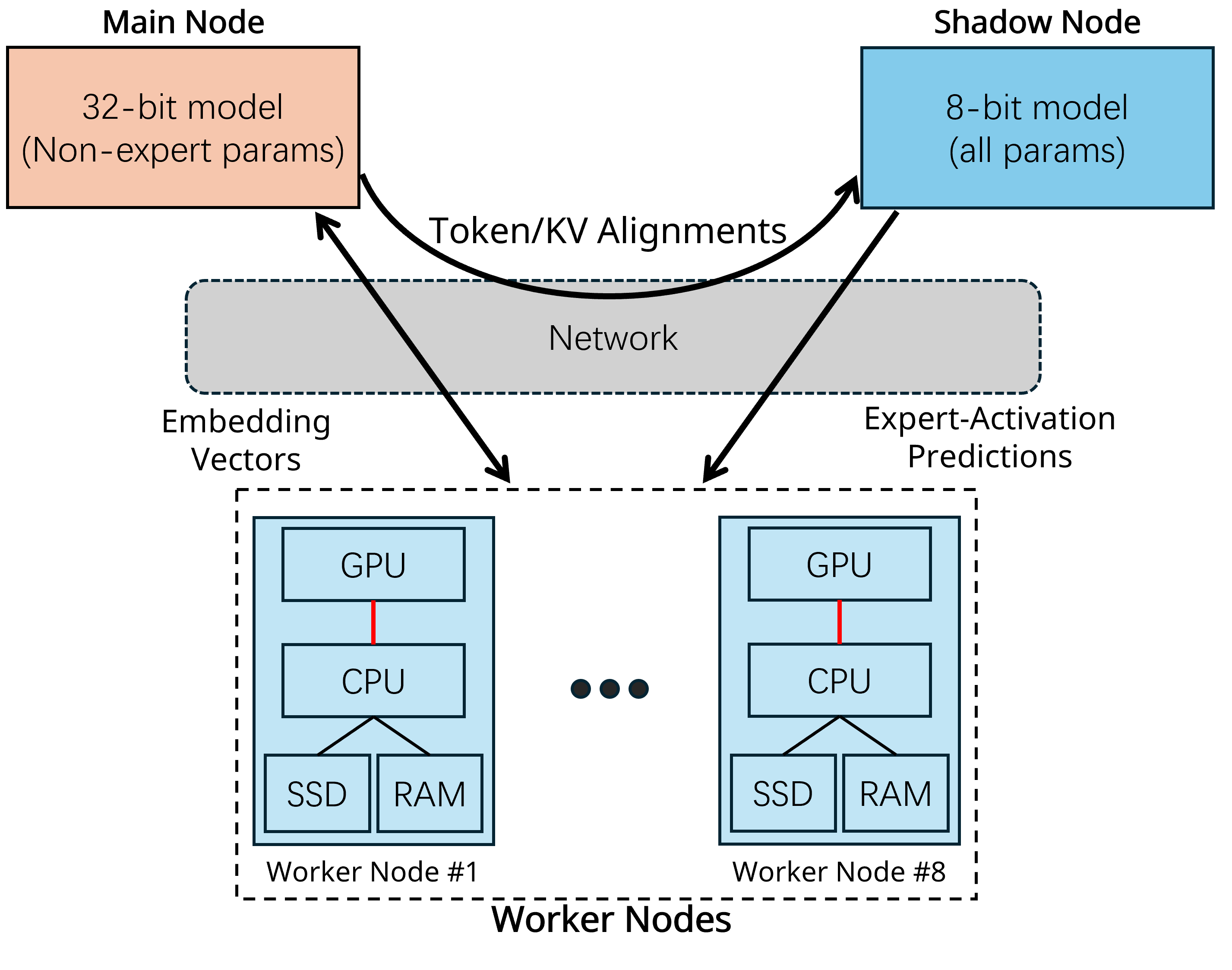}
    \caption{Architecture of OD-MoE. The example here shows the ten-node testbed we have developed, which includes eight worker nodes, one main node, and one shadow node.}
    \label{fig:1_system_design}
\end{figure}

Fig. \ref{fig:1_system_design} illustrates the system architecture of OD-MoE using the ten-node testbed we developed. OD-MoE consists of three types of nodes: 1) a \textbf{main node} hosting all non-expert components in the MoE model, including attention networks, gating networks, normalization networks, and others; 2) a \textbf{shadow node} runs SEP in parallel with the main node to predict the expert activation and notify the worker nodes accordingly; and 3) \textbf{worker nodes} that dynamically load experts into the GPU based on prediction and execute the associated expert computations.

\begin{figure*}
    \centering
    \includegraphics[width=0.925\linewidth]{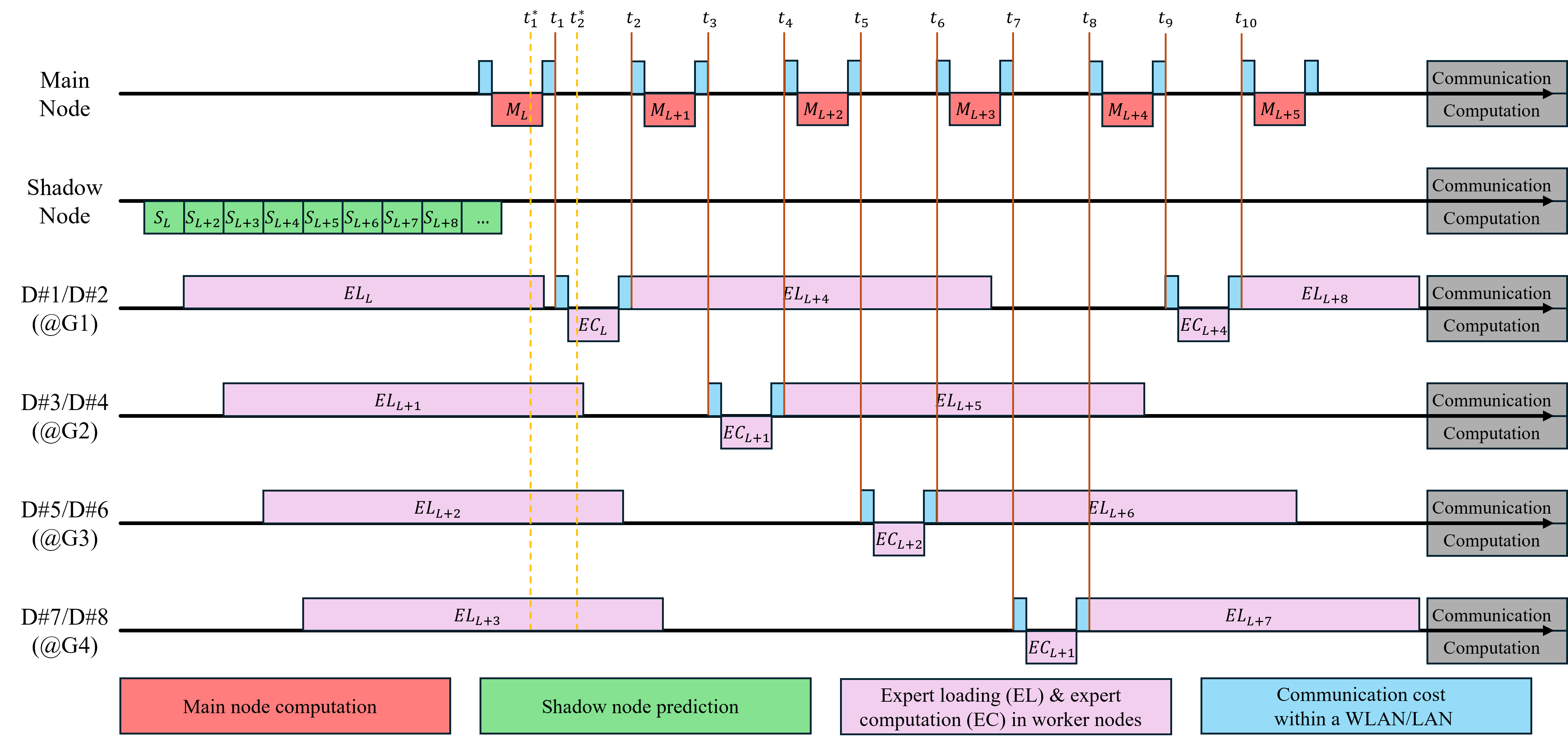}
    \caption{Timing diagram of OD-MoE illustrating the round-robin scheduling scheme therein. Main node computation for layer $l$ is denoted by $M_l$. Shadow node computation for layer $l$ is denoted by $S_l$. Expert loading and expert computation for layer $l$ are denoted by $EL_l$ and $EC_l$}
    \label{fig:2_timeline}
\end{figure*}

Key methodologies in OD-MoE include worker-node grouping and round-robin scheduling. These concepts are introduced alongside the timing diagram shown in Fig. \ref{fig:2_timeline}.

\textbf{Worker-Node Grouping.} With the several-layer-ahead predictions provided by the shadow node, OD-MoE employs a “group-and-schedule” strategy to orchestrate the workload of distributed workers. This strategy aims to mitigate the GPU I/O pressure by parallelizing expert activations (handled by one group of devices) and on-demand expert loadings (handled by other groups of devices). Specifically, worker nodes are divided into $N_W/G$ groups, where $N_W$ is the number of workers (eight in our testbed) and $G$ is the group size. All workers within a group either load or execute experts in parallel for a particular MoE layer in a synchronous manner. In our implementation, $G=2$ since a top-2 activation policy is applied in the Mixtral-8x7B model. Each expert is loaded to only one worker node. This one-to-one assignment of experts to workers prevents uneven expert distributions that could overload specific worker nodes. 

\textbf{Round-Robin Scheduling.} As illustrated in Fig. \ref{fig:2_timeline}, MoE inference for successive layers is scheduled across different groups of devices in a round-robin manner. That is,
\begin{enumerate}[a.]
    \item While devices in Group 1 perform expert computation $EC_l$ from $t_1$ to $t_2$, devices in Group 2, Group 3, and Group 4 simultaneously perform expert loading for future layers $EL_{l+1}$, $EL_{l+2}$, and $EL_{l+3}$ respectively, in a staggered manner. 
    \item Once $EC_l$ completes at $t_2$, devices in Group 1 (Devices \#1 and \#2) shift to expert loading task $EL_{l+4}$ to get ready for future expert computation $EC_{l+4}$.
    \item Devices \#1 and \#2 send embedding generated in $EC_l$ to the main node, and the main node receives the embedding after a communication delay (the tiny blue segments in Fig. \ref{fig:2_timeline}). The main node performs its non-expert computations $M_{l+1}$, including attention, gating, and normalization.
    \item Once $M_{l+1}$ completes at $t_3$, the main node sends the embedding vector to Group 2 (Devices \#3 and \#4) for processing at the next layer, with a similar communication delay.
    \item Operations after $t_3$ mirror Step $a$ to Step $d$ above. The pipeline repeats until computations for all layers are done.
\end{enumerate}

We highlight several important technical details in the pipeline. First, the communication time in Steps $c$ and $d$ in the above includes the time spent on the following components: 1) converting the embedding vector into a packet, 2) delivering the packet over the network (e.g., over WLAN/LAN), 3) converting the packet back into the embedding vector for GPU computation.

Second, we note that a gating network is part of the non-expert parameters stored in the main node. During a main-node computation task, such as $M_{l+1}$ in Step $d$ above, the gating network is activated as in a standard MoE system to determine expert routing for the current layer. If the routing results agree with the predictions generated by the shadow node, no additional expert loading is necessary (e.g., by $t_3$, all predicted experts will have already been loaded into Devices \#3 and \#4). In rare instances where prediction errors occur, the pipeline falls back to the conventional approach, where the expert computation task waits for the completion of expert reloading. However, since such cases are infrequent, their impact on the system’s overall inference speed is minimal.

Third, as shown in Fig. \ref{fig:2_timeline}: 1) when the main node performs its computation task, up to $N_W$ devices (i.e., all devices) can work concurrently to load experts (see $t_1^*$ in Fig. \ref{fig:2_timeline}); 2) when a group of worker nodes is busy with expert computations, the remaining $N_W-k$ devices load experts concurrently (see $t_2^*$ in Fig. \ref{fig:2_timeline}). In essence, the former increases the effective throughput of the system’s CPU-GPU I/O link by $N_W$-fold, while the latter boosts the effective throughput by $(N_W-k)$-fold. For easier identification, critical CPU-GPU links are highlighted in red in Fig. \ref{fig:2_timeline}. The multiplied throughput is a key advantage of OD-MoE over expert-offloading systems with a single node (see Section \ref{Section_II} for related works).

Finally, we provide the timing requirement for fully eliminating the CPU-GPU I/O bottleneck, assuming correct expert-activation prediction. Suppose that the main node requires $t^M$ to complete the main-node computation task (including the communication overhead), i.e., $t^M=t_3-t_2=t_5-t_4=\cdots$.

Similarly, suppose that a worker node requires $t^W$ to complete the expert computation task (also including the communication overhead), i.e., $t^W=t_2-t_1=t_4-t_3=\cdots$.

Then, the maximum allowable duration allowed for the expert loading task without introducing an I/O bottleneck, as shown in Fig. \ref{fig:2_timeline}, is
\begin{equation}
t^{maxload}=Gt^M+(G-1)t^W
\end{equation}

For example, for expert loading task $EL_{l+4}$ starting at $t_2$, there is no compute stall for $EC_{l+4}$ waiting for expert loading to complete if the expert loading completes before $t_9$, i.e., $t^{maxload}(EL_{l+4})=t_9-t_2=4t^M+3t^W$.

In practice, since the parameter size of an expert and the available throughput of a device’s CPU-GPU link are known, we can calculate the required expert loading time and compare it with $t^{maxload}$ to determine whether the system is I/O-bottlenecked.

\subsection{Token and KV Cache Alignments in Shadow Model}\label{Section_III_B}
Model quantization is a well-established technique originally designed to reduce the memory footprint and computational cost of LLM inference, often with only modest degradation in the quality of the LLM’s responses. Consequently, when a quantized model and its full-precision counterpart process identical inputs, their internal states—such as embedding vectors, attention scores, and output logits—are expected to be highly similar. In particular, the expert selection pattern of a quantized model closely mirrors that of the full-precision model for the same input.

\begin{figure}[!h]
\centering
	\subfloat[NF4]{\label{fig:3a}\includegraphics[width = 0.485\textwidth]{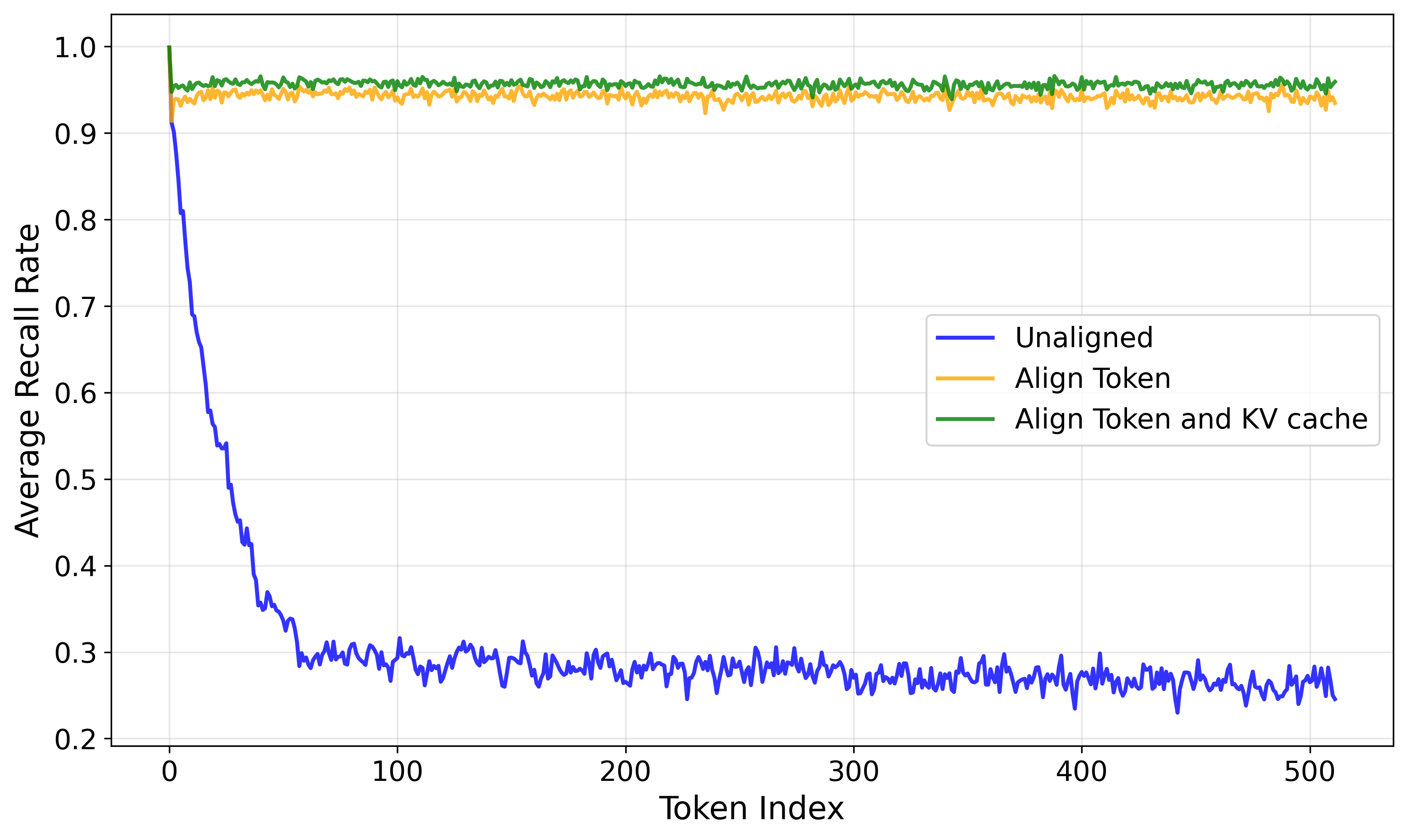}}
    \\
	\subfloat[INT8]{\label{fig:3b}\includegraphics[width = 0.485\textwidth]{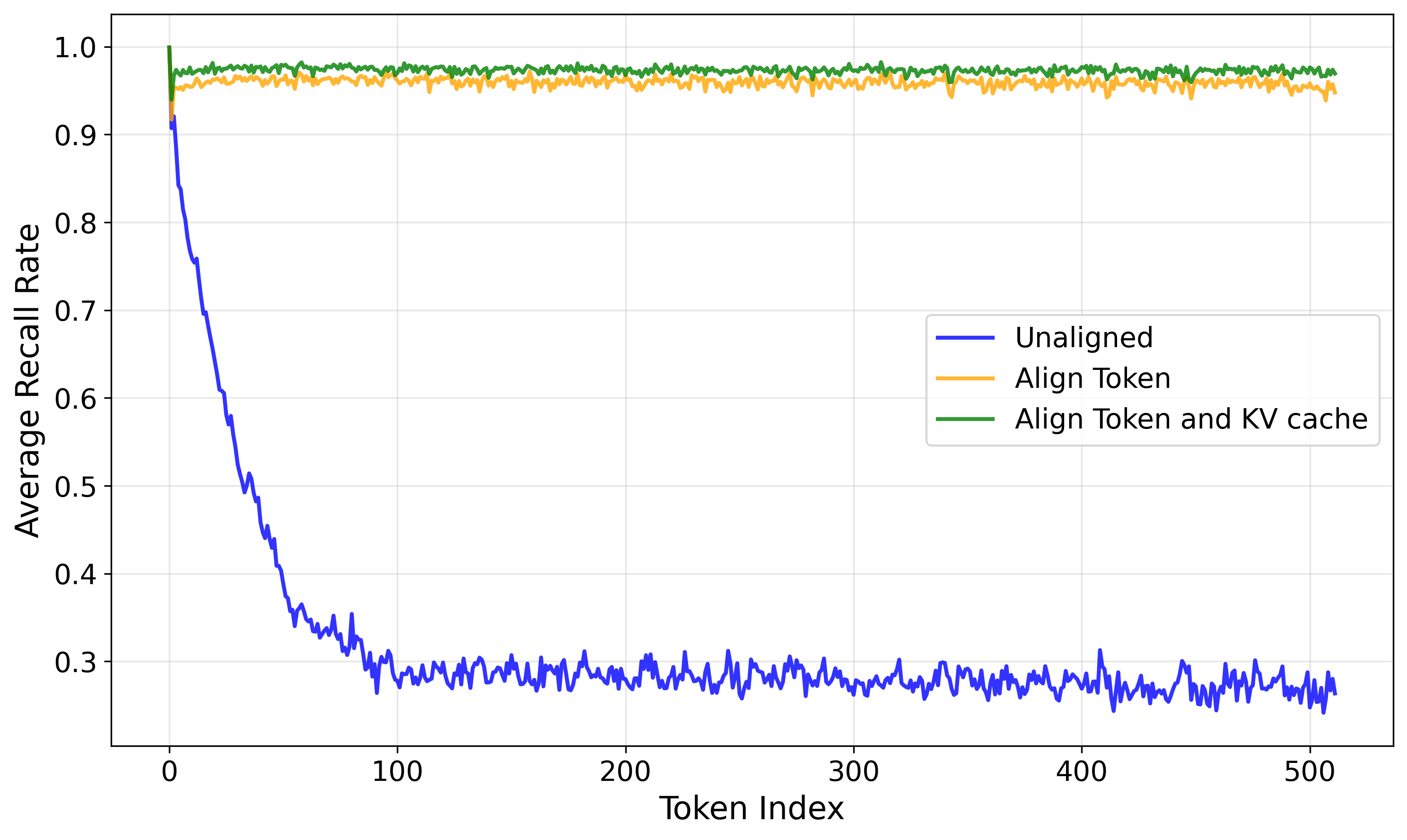}} 
    \\
	\subfloat[FP16]{\label{fig:3c}\includegraphics[width = 0.485\textwidth]{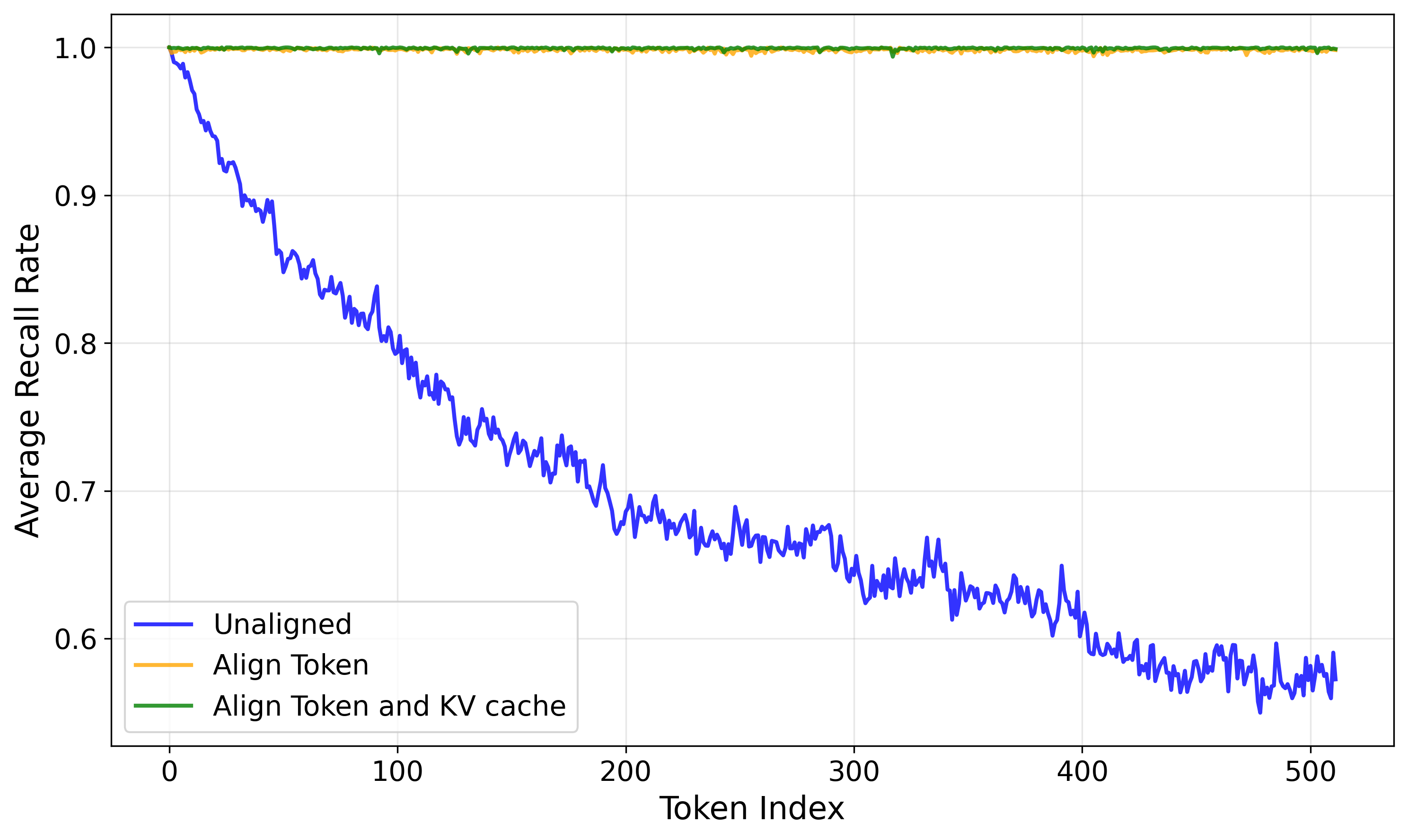}} 
\caption{Expert-selection recall rate versus output token index. Three different quantization schemes (NF4, INT8, and FP16) have been considered for the shadow model, while the original model is realized with a precision of FP32.}
\label{fig:3_quantized_recalls}
\end{figure}

We have experimentally observed this phenomenon during the generation of the first token in the decoding stage. However, as additional tokens are generated autoregressively, the prediction accuracy of SEP deteriorates rapidly (see blue curves in Fig. \ref{fig:3_quantized_recalls}). This degradation is the result of small numerical errors introduced by quantization, which compound during the autoregressive token generation process. The divergence in the shadow model’s expert selection can be attributed to two sources: 1) \textbf{Divergent Generation Paths}: the quantized model may generate a different token than the full-precision model. 2) \textbf{Internal KV-State Drift from Routing Differences}:  differences in expert selections between the quantized and full‑precision models can cause their embeddings to diverge, resulting in discrepancies in their KV cache.

A direct approach to restoring accurate expert-activation prediction in SEP is to align the shadow model’s full KV caches of all attention heads of all decoding layers and tokens with those of the full-precision model to prevent error accumulation. Fig. \ref{fig:3_quantized_recalls} presents experimental results showing the effectiveness of this approach. The results of different alignment set-ups are shown:  1) unaligned shadow model, 2) token aligned after each autoregressive iteration, 3) both token and KV cache aligned after each autoregressive iteration.

Fig. \ref{fig:3_quantized_recalls} shows the prediction accuracy of SEP across various different alignment setups. We follow previous works \cite{tang2024hobbit, zhang2025daop, zhong2024adapmoe} and evaluate prediction accuracy using the recall rate. The recall rate is defined as the number of correctly predicted experts out of all activated experts.

To be precise, consider an MoE model with $L$ layers that employs a top-$k$ activation policy. Let $Q$ be the number of test prompts and $N$ be the maximum number of decoding iterations (i.e., decoded output tokens) in our experiments. We terminate the decoding stage of a prompt when the number of tokens exceeds $N$. Also, the decoding of some of the $Q$ prompts may end before output token $N$. To capture this, let $A(q,n)\in\{0,1\}$ be an indicator for the existence of token $n\in[1,N]$ for prompt $q\in[1,Q]$. For expert-activation predictions, let $c(q,n,l)\in[0,k]$ be the number of correctly predicted experts for layer $l\in[1,L]$, iteration $n$, and prompt $q$.

With the above, the average recall rate for output token n in the decoding stage is defined as 
\begin{equation}
    recall(n)=\sum_{q=1}^Q \sum_{l=1}^L c(q,n,l)A(q,n) \bigg/ kL\sum_{q=1}^Q A(q,n) \label{recall_def}
\end{equation}
and the overall recall rate over the tokens observed is given by
\begin{equation}
    recall=\sum_{n=1}^N \sum_{q=1}^Q \sum_{l=1}^L c(q,n,l)A(q,n) \bigg/ kL\sum_{n=1}^N \sum_{q=1}^Q A(q,n) \label{recall_overall_def}
\end{equation}

In this paper, $k=2$ and $L=32$ for the Mixtral-8x7B model tested. We selected $Q=100$ test problems randomly from the LongWriter dataset \cite{bai2024longwriter} and set $N=512$ for experimental observations. 

Fig. \ref{fig:3_quantized_recalls} presents the recall rate versus token index for the sequence of tokens generated during the decoding stage. As shown, both KV cache alignment and token alignment play a vital role in ensuring the prediction accuracy. By synchronizing tokens and the KV cache for each autoregressive iteration during inference, SEP yields approximately 99.94\%, 97.34\%, and 95.67\% recall rates given by (\ref{recall_overall_def}), with FP16, INT8, and NF4 quantization in the shadow model, respectively. In Section \ref{Section_IV}, we will show that even the NF4-quantized model beats state-of-the-art baselines.

\begin{figure}[htbp]
    \centering
    \includegraphics[width=\linewidth]{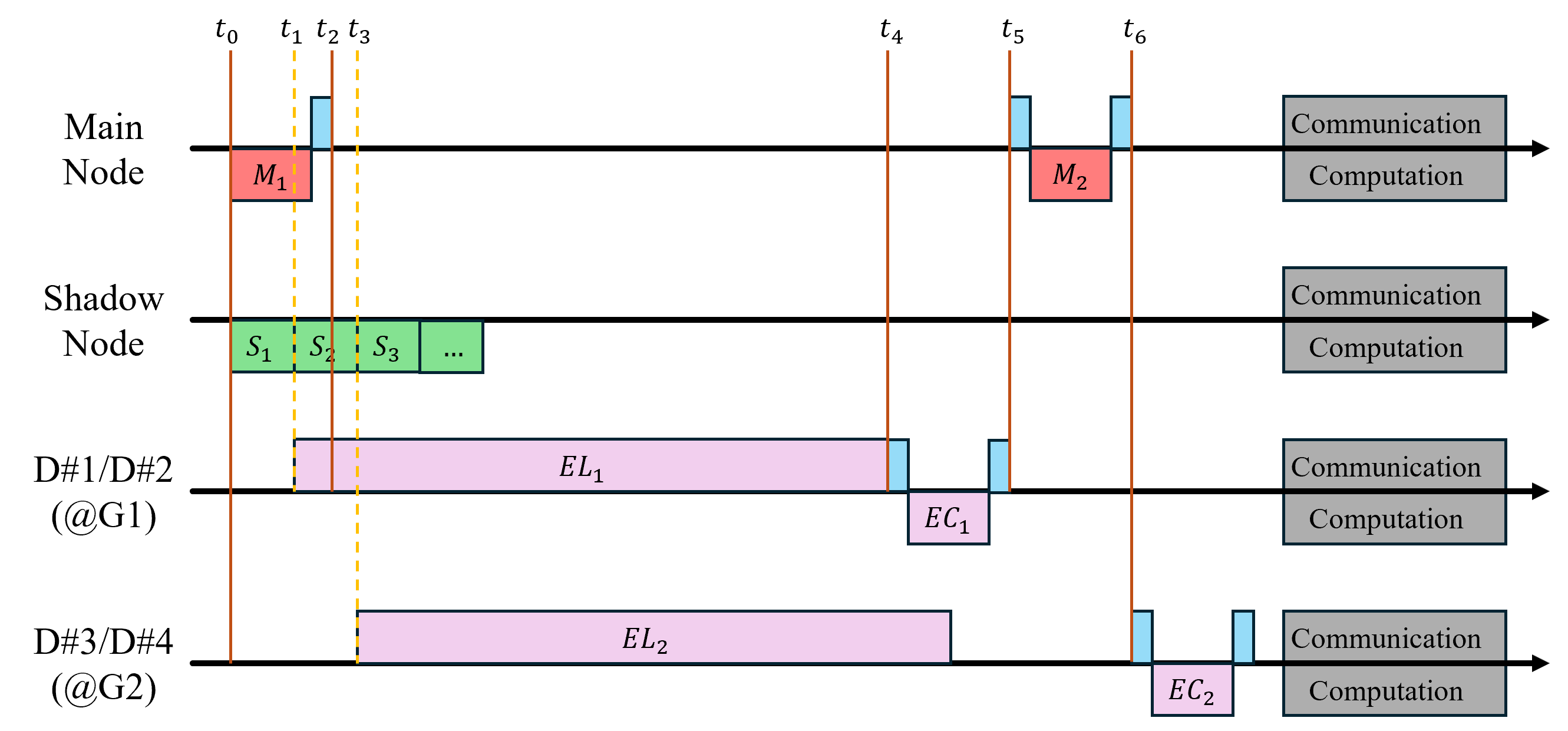}
    \caption{Illustrative timing diagram of OD-MoE, without alignment for the shadow model.}
    \label{fig:4_timeline_without_align}
\end{figure}

\begin{figure}
    \centering
    \includegraphics[width=\linewidth]{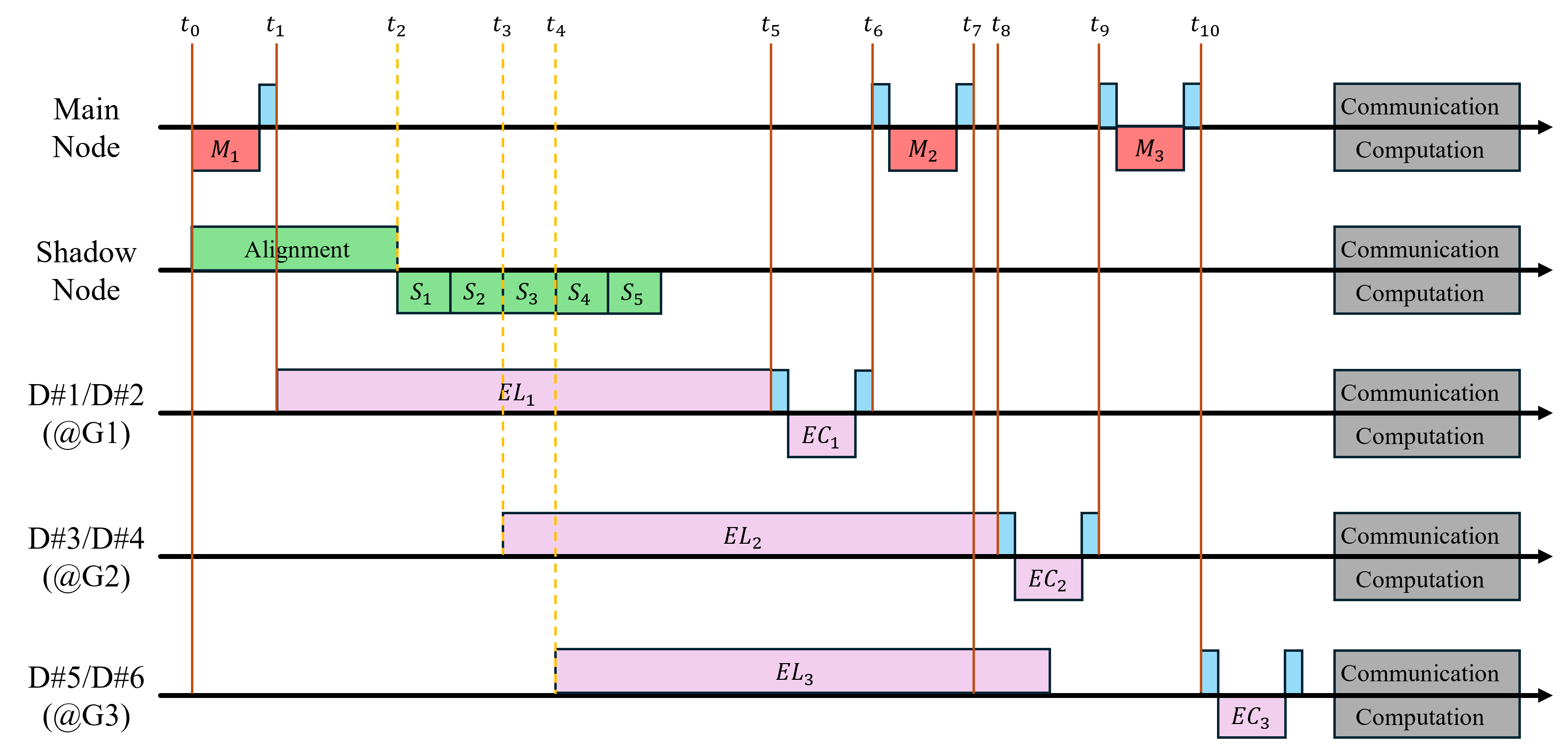}
    \caption{Illustrative timing diagram of OD-MoE, with alignment for the shadow model.}
    \label{fig:5_timeline_with_align}
\end{figure}

The next issue we need to consider is the implementation of KV cache and token alignment. Frequent KV cache and token alignments introduce additional latency referred to as the “late-departure cost”. To better understand that timing cost, we start with a normal timing diagram in Fig. \ref{fig:4_timeline_without_align} to illustrate how the shadow model starts the inference without KV or token alignment. After that, we present the timing diagram in Fig. \ref{fig:5_timeline_with_align} to aid visualization on how alignments cause the shadow model’s “late departure”, and what cost it introduces.

In Fig. \ref{fig:4_timeline_without_align}, both the shadow node and the main node kick off the inference for the first layer at $t_0$. The shadow node runs faster and provides its routing results as a reference at $t_1$, which triggers the expert loading task $EL_1$ for Devices \#1 and \#2. Although $M_1$ ends shortly at $t_2$, the associated expert computation task $EC_1$ has to wait for the completion of $EL_1$ at $t_4$. As indicated in Fig. \ref{fig:4_timeline_without_align}, an I/O bottleneck appears during the inference of the first MoE layer. Subsequent layers avoid I/O stalls. Task $EL_2$ starts on Devices \#3 and \#4 as early as $t_3$. By the time task $M_2$ ends, expert needed for layer 2 have been loaded into the two workers so that the task $EC_2$ does not need to wait as $EC_1$ does.

To reduce the negative impact of alignments on decoding latency (see our later discussions), we could align the tokens and KVs once every few autoregression iterations, at the cost of a lower recall rate. We note that the alignment periods for tokens and KVs could be different. The details of the impacts of alignment periods on decoding speed from experimental results will be presented Section \ref{Section_IV}. Here in Section 3, we explain how alignments can lead to increased decoding latency by timing diagrams. We also present some results on the impacts of alignment frequencies on the recall rate.

Given that 1) the most up-to-date KV cache and tokens of the original full-precision model are available only when the current autoregressive iteration ends, and 2) the main node moves on to the next iteration immediately after the completion of the current iteration, we can use the beginning part of the new iteration to align the shadow model with the latest KV cache and token generated by the full model. That leads to a shadow model’s “late departure” problem as illustrated in Fig. \ref{fig:5_timeline_with_align}.

In Fig. \ref{fig:5_timeline_with_align}, expert loading task $EL_1$ relies on the routing results of the main model available at $t_1$. The routing prediction for layer 2 is not available until the shadow model catches up at $t_3$. Although $EL_2$ starts immediately on Devices \#3 and \#4 at $t_3$, the loading task has not yet completed by the time $M_2$ ends at $t_7$. Thus, the I/O bottleneck persists in layer 2, keeping expert computation task $EC_2$ waiting until $EL_2$ ends at $t_8$.

\begin{figure}[htbp]
    \centering
    \includegraphics[width=0.95\linewidth]{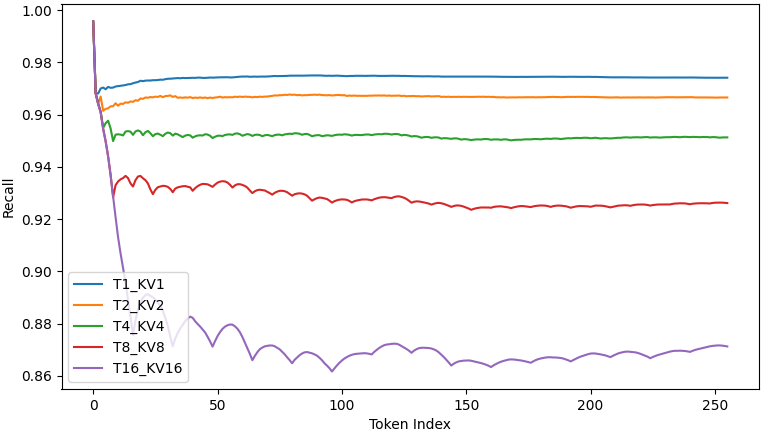}
    \caption{Average expert activation prediction accuracies for token and KV cache alignment intervals of 1/2/4/8/16. In the legend, $T_i\_KV_j$ denotes token alignment period of $i$ and KV alignment period of $j$. An INT8 quantized shadow model is used, reaching over 97.34\% prediction accuracy with the $T_1\_KV_1$ setup.}
    \label{fig:6_recall_of_align_periods}
\end{figure}

Comparing Fig. \ref{fig:4_timeline_without_align} and Fig. \ref{fig:5_timeline_with_align}, we see that the late departure of the shadow model, caused by the alignment operation, prolongs the I/O bottleneck, thus adding to the decoding latency. Fig. \ref{fig:6_recall_of_align_periods} presents the experimental results about the relationship between recall rate and token/KV alignment periods.

We see that there is a clear trade-off we need to consider between the “late-departure cost” and the reduced accuracy in terms of expert-activation prediction. Section \ref{Section_IV} experimentally investigates this balancing required to obtain the optimal setup for the alignment operations.

\subsection{Prefilling-Stage Batched Processing}\label{Section_III_C}
In prefilling, the entire input prompt is processed in a single batch to create the initial attention KV cache, which is used to generate the first output token in the decoding stage. With batched execution, token embeddings routed to the same expert can be grouped and computed via a larger matrix multiplication. Expert-activation prediction and on-demand loading are not performed during prefilling due to the high number of expert activations in batched operations. Even if attempted, prediction offers minimal practical benefit, as all experts are likely to be invoked during batching\footnote{This claim is justified by our experimental results in Section \ref{Section_IV}. In the prefilling stage, short inputs with 16 tokens activates 7.6 experts out of total 8 experts on average, while long inputs with 128 tokens activate all 8 experts with 99.8\% probability.} and must be loaded in any case. Instead, we load the eight experts for each layer onto eight worker nodes in parallel (i.e., each worker handles one expert of every layer). Once loaded, embeddings are grouped by their desired experts, and batched expert activations are executed across the eight workers in parallel.

Our system splits a large batch into multiple mini-batches to improve GPU utilization on worker nodes. This design is motivated by a unique characteristic of edge scenarios: the LAN/WLAN link connecting the main node and worker nodes typically has lower throughput compared to the high-bandwidth, low-latency interconnect fabrics found in data centers. As a result, the communication cost for transmitting batched embeddings (from the main node to a worker) is non-negligible.

\begin{figure}[htbp]
\centering
	\subfloat[Prefilling with one single large batch]{\label{fig:7a}\includegraphics[width = 0.49\textwidth]{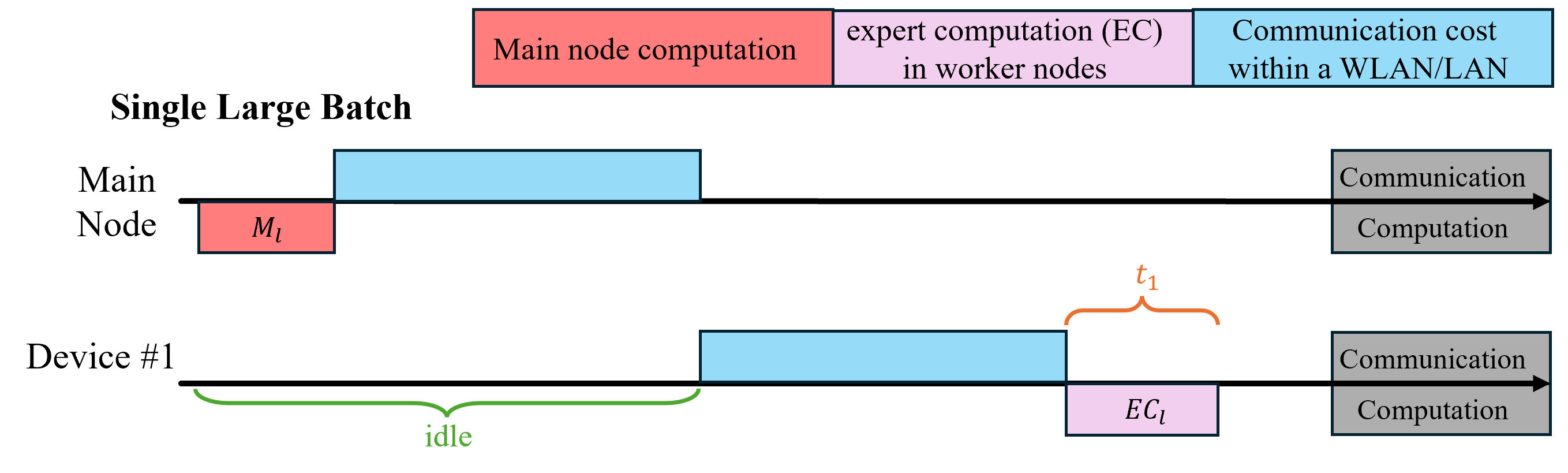}} \\
	\subfloat[Prefilling with multiple mini-batches]{\label{fig:7b}\includegraphics[width = 0.49\textwidth]{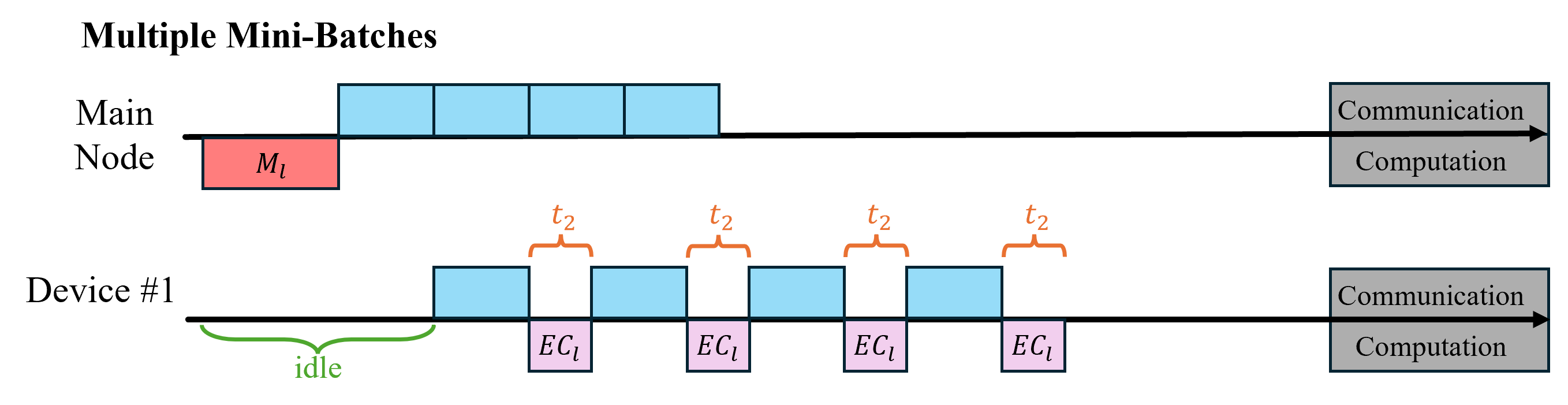}} \\
\caption{Illustrative timing diagrams for batched expert computations during the prefilling stage: (a) The main node transmits a single large batch to a worker. (b) The main node splits a large batch into mini-batches for transmission to a worker. We present the timing diagram for Device \#1 only, since prefilling behaviors are similar across worker nodes.}
\label{fig:7_chunked_prefill}
\end{figure}

As illustrated in Fig. \ref{fig:7a}, without mini-batching, the worker remains idle while the embedding vectors are being transmitted over the edge network, leading to reduced GPU utilization efficiency. In contrast, Fig. \ref{fig:7b} shows that dividing a large batch into multiple mini-batches reduces the worker's idle time by allowing it to begin expert computation as soon as the first mini-batch is received. In other words, mini-batching enables pipelined processing.

Although the total computation time for the multiple mini-batches (i.e., $4t_2$ in Fig. \ref{fig:7b}) is larger than that of a single large batch (i.e., $t_1$ in Fig. \ref{fig:7a}) given the computation efficiency of a larger batch, mini-batched operation in Fig. \ref{fig:7b} saves more time owing to the pipelined computation, eventually results in lower prefilling latency in general.

\section{Experiments}\label{Section_IV}
This section begins with an overview of the experimental setup in Subsection \ref{Section_IV_A}. Subsection \ref{Section_IV_B} investigates the alignment process through ablation studies and parameter optimizations. Subsection \ref{Section_IV_C} compares the prediction accuracy of SEP with state-of-the-art baselines. Subsection \ref{Section_IV_D} benchmarks OD-MoE against existing MoE inference frameworks.

\subsection{Experimental Setup}\label{Section_IV_A}
\textbf{Base Model and Configurations.} We use Mixtral-8x7B \cite{jiang2024mixtral} as the base model and employ the greedy decoding policy \cite{song2025good} for token generation. Greedy decoding selects the maximum-likelihood tokens in the output logits (i.e., not probabilistic sampling), ensuring better reproducibility.

\textbf{Hardware Setup.} Our testbed, as shown in Fig. \ref{fig:1_system_design}, consists of ten nodes. The main node runs non-experts on a single NVIDIA RTX 3090 GPU, while the shadow node runs a quantized Mixtral-8x7B model (INT8, unless stated otherwise) on two NVIDIA RTX 3090 GPUs. The remaining eight worker nodes each have an NVIDIA RTX 3090 GPU. All nodes use AMD R7-7700 CPUs. Worker nodes have 192 GB of DRAM, while the main and shadow nodes each have 128 GB. The nodes are connected via a 1Gbps Ethernet LAN.

\textbf{Speed Evaluation Metrics.} Following the classic setup \cite{mitzenmacher2025queueing}, we evaluate the inference speed of the decoding stage by its throughput, i.e., the average number of tokens generated per second during the decoding. The speed of the prefilling stage is measured by Time-To-First-Token (TTFT), and the overall speed is represented by the throughput for the whole inference process including the prefilling and decoding stages.

\textbf{Test Setup (Inference Speed).} Subsections \ref{Section_IV_B} and \ref{Section_IV_D} involve the evaluation of model inference speed. To facilitate reproduceable benchmarking across existing approaches, we follow the setup in previous works for speed test. Specifically, we inherit the test dataset from HOBBIT \cite{tang2024hobbit}. The test dataset contains a subset of 60 high-quality samples from the well-known Alpaca dataset \cite{alpaca}, in which 30 of them have 16 input tokens while the remaining 30 samples have 128 input tokens.

All 60 prompts can generate at least 256 output tokens if left to run through their decoding iterations. However, we limit the number of iterations (output tokens) to either 64 or 256 in compiling our test results. The reason for doing so is to ensure fair comparisons between different schemes -- ours as well as others -- so that all schemes can be compared on the basis of the same number of decoding iterations (output tokens). Different inference frameworks may result in different output lengths if left to run, even with the same prompt, but all of them can also generate at least 256 output tokens.

The above setup results in four evaluation cases represented as (input length, output length) tuples: (16, 64), (16, 256), (128, 64), and (128, 256). For Subsection \ref{Section_IV_B}, we consider the (16, 256) configuration only, while we consider all the four input/output configurations in Subsection \ref{Section_IV_D} for comprehensive benchmarking previous approaches.

\subsection{Experimental Investigations of Alignment Process}\label{Section_IV_B}
This subsection investigates how SEP’s alignment setups impact OD-MoE’s decoding speed using the (16, 256) configuration. We conduct ablation studies to isolate the contributions of KV cache alignment and token alignment to decoding speed. The six cases are as follows:
\begin{enumerate}
    \item Shadow node enabled; token and KV cache alignments performed every iteration.
    \item Shadow node enabled; only token alignment performed every iteration.
    \item Shadow node enabled; only KV cache alignment performed every iteration.
    \item Shadow node enabled; neither alignment operation performed.
    \item Shadow node removed; worker nodes prefetch experts randomly.
    \item Shadow node removed; worker nodes load experts only after receiving routing results from the main node.
\end{enumerate}

To provide context regarding the data volume in the alignment process, we note that the full precision KV cache size on the main node is approximately 8 KB per token per decoding layer. In one alignment run, the main node sends all KV cache corresponding to the newly generated token to the shadow node, resulting in 256 KB (8 KB per token per layer times 32 decoding layers) of data transmission. The aligned token size is negligible (only a few bytes). The transmitted embedding size between the main node and a worker node is approximately 16 KB per token per layer. All data is transmitted over a 1 Gbps Ethernet connection.

\begin{figure}
    \centering
    \includegraphics[width=0.98\linewidth]{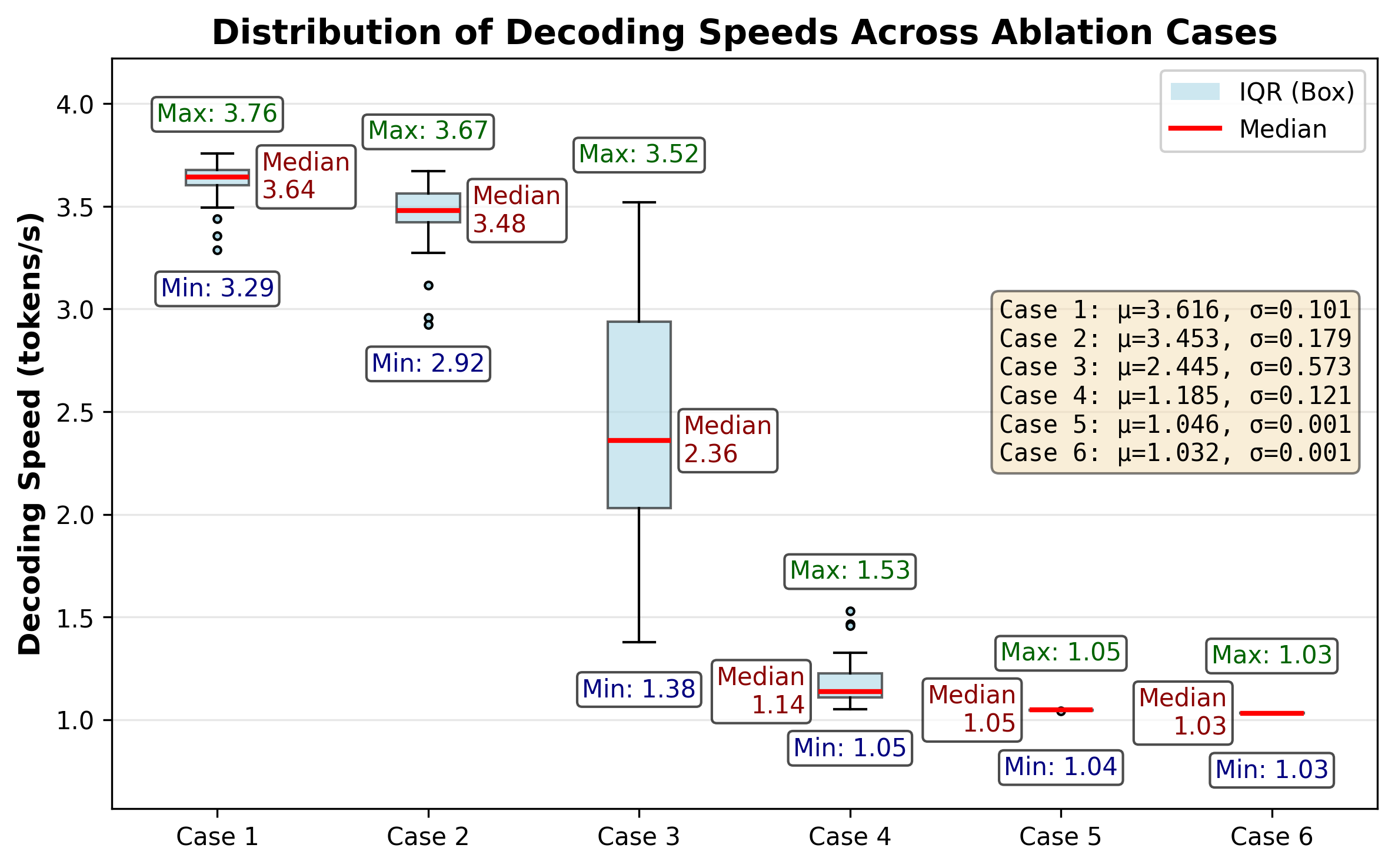}
    \caption{Average decoding speed (tokens/s) for different ablation setups. }
    \label{fig:8_alignment_ablation}
\end{figure}

Fig. \ref{fig:8_alignment_ablation} presents the results. The results of Cases 1–3 indicate that token alignment is more critical than KV‑cache alignment – larger gap between Case 1 and 3 than the gap between Case 1 and 2. This is consistent with Fig. \ref{fig:3_quantized_recalls}, where removing token alignment causes a larger drop in prediction accuracy. 

On the other hand, the smaller yet noticeable gap between Cases 1 and 2 highlights that KV cache alignment is still essential for optimizing decoding speed. Despite the mean value in decoding speed, the increased standard deviation from Case 1 to Case 2 serves as another evidence in Fig. \ref{fig:8_alignment_ablation} that supports the necessity of KV cache alignment. Moreover, we note that experiments in Fig. \ref{fig:8_alignment_ablation} are conducted with 256 output tokens only – with more decoding iterations, removing the KV cache alignment as in Case 2 is expected to result in more severe deterioration in the decoding speed because the KV discrepancy accumulates as the number of token increase. 

The drop in decoding speed from Case 4 to Case 5 also aligns with Fig. \ref{fig:3_quantized_recalls}. In Case 4, the recall of the INT8‑quantized model without KV or token alignment (i.e., the blue curve in Fig. \ref{fig:3b}) decreases from 100\% to around 30\% as the token index increases from 1 to 256, yielding an average recall of \textasciitilde45\%, whereas an easy calculation shows that the recall in Case 5 is only \textasciitilde25\%. The higher recall in Case 4 explains its better decoding speed in comparison to Case 5. 

Finally, the small additional drop in decoding speed from Case 5 to Case 6 suggests that random preloading offers limited benefit. As a general takeaway, the monotonic decrease from Case 1 to Case 6 demonstrates the effectiveness of each component in our MoE inference framework.

\begin{figure}
    \centering
    \includegraphics[width=0.83\linewidth]{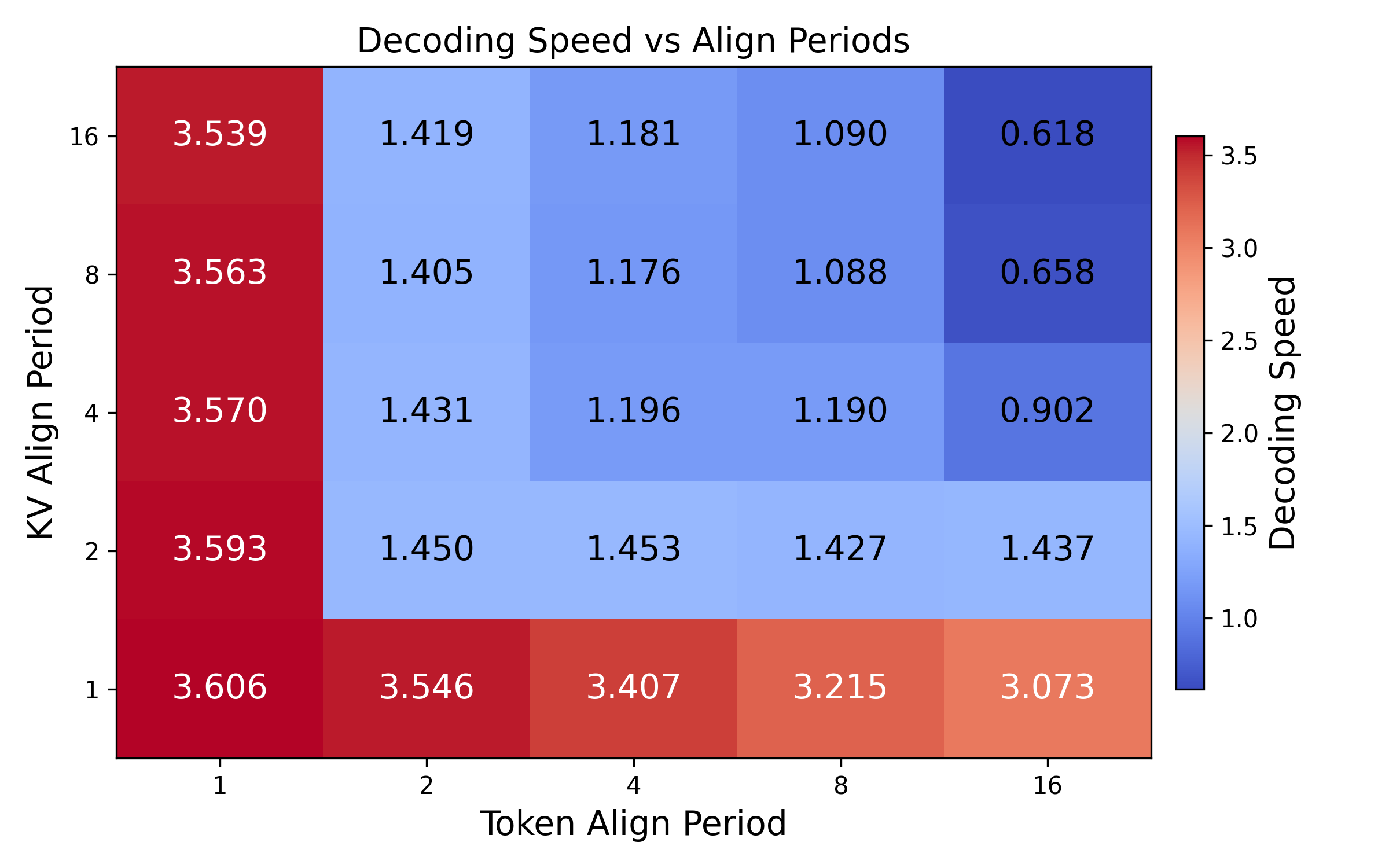}
    \caption{Average decoding speed (tokens/s) of OD‑MoE under different token and KV‑cache alignment periods.}
    \label{fig:9_parameter_setting}
\end{figure}

Following the ablation study, we next study the optimal alignment periods for tokens and KV cache. Recall from Section \ref{Section_III_B} that alignment introduces a trade‑off between late‑departure latency and reduced prediction error. Fig. \ref{fig:9_parameter_setting} reports the decoding speeds for the alignment periods of 1/2/4/8/16 for both token and KV‑cache. As shown, the best decoding speed occurs when both token and KV‑cache alignments use a period of one. In subsequent experiments comparing OD‑MoE with baselines (see Subsection \ref{Section_IV_D}), we adopt this configuration.

\begin{figure}
    \centering
    \includegraphics[width=0.92\linewidth]{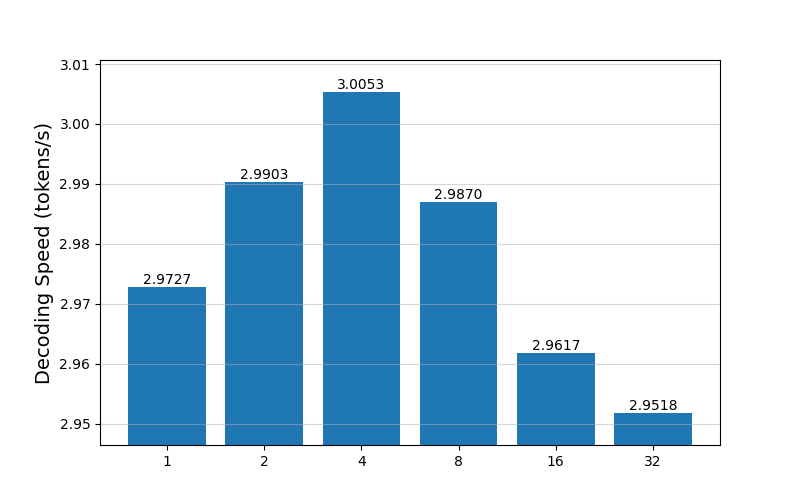}
    \caption{Average decoding speed (tokens/s) of OD‑MoE with worker‑node GPUs replaced by RTX 3080s. The token‑alignment period is fixed at 1; the KV‑cache alignment period varies over \{1, 2, 4, 8, 16, 32\}.}
    \label{fig:10_parameter_setting_3080}
\end{figure}

We emphasize that although the highest speed in Fig. \ref{fig:8_alignment_ablation} is achieved with a simple configuration (i.e., aligning tokens and KV cache every decoding iteration), this does not mean the trade‑off posed by the late-departure issue discussed in Section \ref{Section_III_C} is not important in general. The results in Fig. \ref{fig:9_parameter_setting} only indicate that reducing prediction error has a much larger impact on speed than lowering late‑departure latency with the experimental hardware setup. In general, the optimal trade‑off depends on both computation time (expert activation) and communication time (expert loading) on the worker node. For example, replacing the worker‑node GPUs with RTX 3080s (see Fig. \ref{fig:10_parameter_setting_3080}) shifts the optimum: the best speed is obtained with a KV‑cache alignment period of four and a token‑alignment period of one.

\subsection{Baseline Comparison: Expert-Activation Prediction}\label{Section_IV_C}
Table \ref{tab:EA_prediction} lists the performance results of some previously reported expert-activation prediction schemes. Prediction accuracy can be evaluated using two metrics: recall rate and cache-hit rate (the probability of having the required expert in the GPU cache). 

\begin{table}[h!]
\centering
\caption{Baseline Comparison in expert-activation prediction.}
\label{tab:EA_prediction}
\begin{threeparttable}
\begin{tabularx}{\columnwidth}{l *{5}{>{\centering\arraybackslash}X}}
\toprule
                 & \thead{Adap} & \thead{DAOP} & \thead{Hobt} & \thead{MxOf} & \thead{fMoE} \\
\midrule
Recall Rate      & 0.86    & 0.84 & 0.91   & /                  & / \\
Cache-hit Rate   & /       & /    & /      & \textasciitilde0.80  & <0.85 \\
\bottomrule
\end{tabularx}
\begin{tablenotes}
\scriptsize
\item \textbf{Note}: Adap = AdapMoE, Hobt = HOBBIT, MxOf = Mixtral Offloading
\end{tablenotes}
\end{threeparttable}
\end{table}

AdapMoE \cite{zhong2024adapmoe} and DAOP \cite{zhang2025daop} achieve recall rates of 86\% and 84\%, respectively. HOBBIT \cite{tang2024hobbit} is more sophisticated – it allows predictions across multiple layers. Specifically, HOBBIT makes predictions up to four layers ahead, and it reports a recall rate of 91\% on average for predicted layers.

Mixtral Offloading \cite{eliseev2023fast} and fMoE \cite{yu2025fmoe} evaluate their predictor by cache-hit rate. As these two projects are either partly open source or closed source, we cannot reproduce their prediction algorithms. However, based on figures in their technical reports, we can roughly estimate their cache-hit rate: the Fig. 2 in \cite{eliseev2023fast} indicates that its cache-hit rate is around 80\% when applying the same top-2 activation policy as in this paper, while the Fig. 12 in \cite{yu2025fmoe} shows that fMoE’s cache-hit rate is no larger than 85\%.

As OD-MoE follows the cache-free design (i.e., only the predicted expert is loaded into the GPU), the cache-hit rate is equivalent to its recall rate. With the test setup described in Section \ref{Section_III_B}, experimental results presented in Fig. \ref{fig:3_quantized_recalls} highlight that SEP achieves 99.94\%, 97.34\%, and 95.67\% recall rates when realized with FP16, INT8, and NF4 quantization, respectively. These experimental results, even with the NF4 quantization, clearly outperform existing methods in both prediction recall rate and cache-hit rate\footnote{Aside from the works discussed here, we do not benchmark against other works mentioned in Section \ref{Section_II}, as they are either partially open source and do not provide the predictor implementation, or they require specialized mobile devices for system deployment. Moreover, their technical reports do not provide direct information on prediction accuracies or cache-hit rates.}.

\subsection{Baseline Comparison: Inference Performance}\label{Section_IV_D}
This subsection benchmarks OD-MoE against existing methods based on three metrics:
\begin{enumerate}
    \item \textbf{Model Inference Speed}: This includes the speeds during the prefilling stage, the decoding stage, and the overall inference process.
    \item \textbf{GPU Memory}: The total GPU memory required by the inference system.
    \item \textbf{Model Performance}: Evaluation on multiple classic LLM benchmarks to measure the quality of the model’s answers. 
\end{enumerate}

We compare OD-MoE against two classes of baselines: 1) full‑precision inference engines with all experts pre‑cached, and 2) representative expert‑offloading systems. The former, which includes HuggingFace Transformers \cite{wolf2020transformers} and llama.cpp \cite{gerganov2023ggerganov}, serves as data‑center references where the MoE model under study (i.e., Mixtral‑8x7B) does not requires expert offloading. The latter includes Mixtral‑Offloading \cite{eliseev2023fast}, MoE‑Infinity \cite{xue2024moe}, HOBBIT \cite{tang2024hobbit}, and AdapMoE \cite{zhong2024adapmoe} – existing open‑source systems discussed previously.

To deploy the full-precision Mixtral-8x7B through the Transformers engine, GPU memory of at least equivalent to eight RTX 3090 GPUs is required. Therefore we reproduce the baseline methods on a GPU server with 1) eight NVIDIA RTX 3090 GPUs, 2) AMD EPYC 7K62 CPU, and 3) 512G DRAM. Note that the offloading baselines do not support distributed computing and thus only leverage one GPU on the server. This hardware setup ensures faithful reproductions of baseline methods and fair comparisons with our system.

Our experimental results are summarized in Table \ref{tab:benchmark}. We first look at Part (i) of Table \ref{tab:benchmark}. For \textbf{\textit{decoding throughput}} – the primary focus of this paper – the Transformers engine with all experts cached achieves 4.8900 tokens/s on average (averaged across four input/output configurations). Note that we do not consider batched decoding here to align with the decoding setup of previous studies \cite{eliseev2023fast, xue2024moe, tang2024hobbit, zhong2024adapmoe} for fair comparisons. OD‑MoE achieves 3.6925 tokens/s on average, retaining 75.51\% of the Transformers engine’s speed. Compared with llama.cpp (0.8225 tokens/s on average), which runs on CPUs and caches all expert parameters in DRAM, OD‑MoE is 4.49× faster owing to GPU acceleration. Against expert‑offloading baselines, OD‑MoE outperforms Mixtral‑Offloading (2.2375 tokens/s), MoE Infinity (0.6875 tokens/s), HOBBIT (0.7850 tokens/s), and AdapMoE (3.1300 tokens/s) by 1.65×, 5.37×, 4.70×, and 1.18×, respectively.

We next examine \textbf{\textit{prefilling speed}}, measured by TTFT. Prefilling latency is strongly dependent on input length: OD‑MoE achieves \textasciitilde1340 ms and \textasciitilde3150 ms TTFT for 16 and 128 input tokens, respectively, with an average of \textasciitilde2244 ms. Compared with offloading baselines, OD‑MoE outperforms MoE Infinity and HOBBIT but trails Mixtral‑Offloading and AdapMoE – likely because these two systems use expert quantization to reduce both loading and computation time. However, as shown in Part (iii) of Table \ref{tab:benchmark}, expert quantization degrades answer quality.

For the output throughput encompassing both prefilling and decoding stages, OD‑MoE averages 3.3700 tokens/s, exceeding Mixtral‑Offloading (2.1725 tokens/s), MoE Infinity (0.6675 tokens/s), HOBBIT (0.7575 tokens/s), and AdapMoE (3.0350 tokens/s) by 1.55×, 5.05×, 4.45×, and 1.11×, respectively.

\begin{table*}[t]
\centering
\caption{Baseline Comparison of inference speed, GPU memory requirement, and model answer quality.}
\label{tab:benchmark}
\resizebox{\textwidth}{!}{%
\begin{tabular}{@{}llrrrrrrr@{}}
\toprule
\textbf{Category} & \textbf{Configuration/Dataset} & \textbf{Mixtral Offloading} & \textbf{MoE-Infinity} & \textbf{HOBBIT} & \textbf{AdapMoE} & \textbf{Transformer} & \textbf{Llama.cpp} & \textbf{OD-MoE} \\
\midrule
\multicolumn{9}{c}{\textbf{(i) Benchmarks in Inference Speed}} \\
\midrule
\multirow{5}{*}{TTFT (ms)} & (16, 64) & 1727.81 & 5521.63 & 5456.37 & 1345.12 & 385.52 & 2025.10 & 1349.78 \\
 & (16, 256) & 1705.15 & 5560.34 & 5467.42 & 1428.23 & 386.86 & 2008.60 & 1340.89 \\
 & (128, 64) & 1967.16 & 5819.49 & 5898.92 & 1343.96 & 447.43 & 6592.81 & 3150.87 \\
 & (128, 256) & 1979.81 & 7573.27 & 5896.04 & 1430.69 & 448.16 & 6527.13 & 3135.85 \\
\cmidrule(lr){2-9}
 & \textbf{Average} & \textbf{1844.9825} & \textbf{6118.6825} & \textbf{5679.6875} & \textbf{1387.0000} & \textbf{416.9925} & \textbf{4288.4100} & \textbf{2244.3475} \\
\midrule
\multirow{5}{*}{\begin{tabular}[c]{@{}c@{}}Decoding Throughput\\ (token/s)\end{tabular}} & (16, 64) & 2.26 & 0.68 & 0.78 & 3.14 & 4.90 & 0.85 & 3.67 \\
 & (16, 256) & 2.26 & 0.68 & 0.79 & 3.15 & 4.87 & 0.82 & 3.65 \\
 & (128, 64) & 2.22 & 0.70 & 0.79 & 3.11 & 4.90 & 0.81 & 3.74 \\
 & (128, 256) & 2.21 & 0.69 & 0.78 & 3.12 & 4.89 & 0.81 & 3.71 \\
\cmidrule(lr){2-9}
 & \textbf{Average} & \textbf{2.2375} & \textbf{0.6875} & \textbf{0.7850} & \textbf{3.1300} & \textbf{4.8900} & \textbf{0.8225} & \textbf{3.6925} \\
\midrule
\multirow{5}{*}{\begin{tabular}[c]{@{}c@{}}Output Throughput\\ (token/s)\end{tabular}} & (16, 64) & 2.16 & 0.66 & 0.74 & 2.99 & 4.84 & 0.84 & 3.41 \\
 & (16, 256) & 2.24 & 0.67 & 0.78 & 2.98 & 4.86 & 0.82 & 3.54 \\
 & (128, 64) & 2.11 & 0.66 & 0.74 & 3.09 & 4.81 & 0.75 & 3.09 \\
 & (128, 256) & 2.18 & 0.68 & 0.77 & 3.08 & 4.86 & 0.78 & 3.44 \\
\cmidrule(lr){2-9}
 & \textbf{Average} & \textbf{2.1725} & \textbf{0.6675} & \textbf{0.7575} & \textbf{3.0350} & \textbf{4.8425} & \textbf{0.7975} & \textbf{3.3700} \\
\midrule[\heavyrulewidth]
\multicolumn{9}{c}{\textbf{(ii) Benchmarks in total GPU Memory Requirement (GB)}} \\
\midrule
\multicolumn{2}{l}{Baseline Models Implemented by Default} & 11 & 21.5 & 22 & 8 & 180 & N/A & 60 \\
\midrule[\heavyrulewidth]
\multicolumn{9}{c}{\textbf{(iii) Benchmarks in Answer Quality}} \\
\midrule
\multirow{2}{*}{General Knowledge} & Hellaswag & 63.50\% & 64.50\% & 62.00\% & 44.50\% & & 76.25\% & \\
 & MMLU & 49.82\% & 51.93\% & 55.26\% & 48.60\% & & 70.34\% & \\
\cmidrule(lr){1-9}
\multirow{2}{*}{Math} & ARC-Challenging & 82.37\% & 77.97\% & 86.10\% & 75.68\% & & 86.44\% & \\
 & GSM8k & 56.00\% & 60.00\% & 35.00\% & 22.00\% & & 64.14\% & \\
\cmidrule(lr){1-9}
\multirow{2}{*}{Reasoning} & WinoGrande & 52.40\% & 56.40\% & 41.20\% & 43.20\% & & 65.67\% & \\
 & BigBenchHard & 58.30\% & 52.20\% & 38.89\% & 43.33\% & & 59.84\% & \\
\cmidrule(lr){1-9}
\multirow{2}{*}{Coding} & Big Code & 8.00\% & 15.00\% & 8.00\% & 0.00\% & & 16.00\% & \\
 & humaneval & 17.68\% & 1.83\% & 1.83\% & 1.54\% & & 24.39\% & \\
\cmidrule(lr){1-9}
Instruction Following & MT-bench-101 & 7.4 / 10 & 7.81 / 10 & 6.67 / 10 & 4.47 / 10 & & 7.83 / 10 & \\
\cmidrule(lr){1-9}
Anti-Hallucination & truthfulQA & 86.50\% & 87.50\% & 73.50\% & 76.50\% & & 89.00\% & \\
\bottomrule
\end{tabular}%
}
\end{table*}

Moving to Part (ii) of Table \ref{tab:benchmark} on \textbf{GPU memory requirement}, OD‑MoE uses only 1/3 of the full‑precision model’s GPU memory. A detailed breakdown is as follows: 7 GB on the main node, 45 GB on the shadow node, and 1 GB per worker across eight workers, resulting in 60GB in total. As llama.cpp targets CPU-oriented computations, its GPU memory usage is nil.

Finally, we look at Part (iii) of Table \ref{tab:benchmark}, answer quality. The evaluation of an LLM’s performance is a broad topic. To narrow down the scope, we select widely used LLM performance benchmarks across six representative categories of interest to the community: 1) General Knowledge \cite{zellers2019hellaswag, hendrycks2020measuring}, 2) Math \cite{clark2018think, cobbe2021training}, 3) Reasoning \cite{sakaguchi2021winogrande,suzgun-etal-2023-challenging}, 4) Coding \cite{zhuo2024bigcodebench, chen2021evaluating}, 5) Instruction Following \cite{bai2024mt}, and 6) Anti-Hallucination \cite{lin2022truthfulqa}. All these benchmarks have open-sourced the evaluation scripts, and most of them present the evaluation results with the percentage of correct/satisfactory answers. The only exception is MT-bench-101 \cite{bai2024mt}, which uses a third-party judge (i.e., following the setup in \cite{bai2024mt}, we use GPT4 as the judge) to score the tested model’s instruction following performance under the setup of multi-turn dialogues. As a full‑precision solution, OD‑MoE matches the answer quality of Transformers and llama.cpp. Meanwhile, as experimental results suggest, OD‑MoE consistently surpasses expert‑offloading baselines across all benchmarks tested within the six categories.

To conclude the discussion of the results, we remark that OD‑MoE, being a full‑precision MoE distributed inference framework with expert offloading, preserves full‑precision answer quality while using comparatively little GPU memory compared with the fully pre-cached solution. OD-MoE demonstrates consistent answer quality advantages over all the offloading baselines which have precision losses, and exhibits superior inference speed comparable to the fully-precached solution and surpassing all offloading baselines. However, the full-precision and competitive inference speed comes with a price of a large GPU memory usage for the shadow node and the network delays absent in the baselines' single-server systems. On the other hand, OD-MoE has the advantage of parallel execution over distributed nodes. More importantly, the worker nodes only require a small GPU memory footprint. The RTX 3090 GPUs in the testbed can be substituted with less powerful entry-level GPUs.

\section{Conclusion}\label{Section_V}
This paper presented \textbf{OD‑MoE}, a cache‑free, on‑demand MoE inference framework designed specifically for memory‑constrained edge environments. The core innovations of OD‑MoE include: \textbf{1) SEP}, an ultra-accurate, multi‑layer lookahead predictor implemented with a quantized shadow model, reaching up to 99.94\% accuracy for expert-activation predictions: and \textbf{2) Grouped workers with round‑robin scheduling}, which overlap expert loading and computation across distributed devices. This parallelization significantly increases effective CPU‑GPU I/O throughput and eliminates the need for long‑term expert caching. 

To ensure stability during long decoding sequences, we introduced an advanced alignment mechanism in SEP, which synchronizes the shadow model’s token generation and KV‑cache with the full‑precision model. On Mixtral‑8×7B, OD‑MoE achieves 75\% of the decoding speed of a full GPU‑cached deployment while using only one-third of the GPU memory. The framework maintains full‑precision answer quality and consistently outperforms representative expert-offloading baselines across multiple evaluation benchmarks. Additionally, OD‑MoE reduces the per‑worker GPU memory footprint to under 1 GB, enabling practical MoE inference on low‑cost edge GPUs and even IoT‑class devices.

{\footnotesize \bibliographystyle{acm}
\bibliography{sample}}


\end{document}